\documentclass[final,5p,times]{elsarticle}
\usepackage{amssymb}
\usepackage{amsfonts}
\usepackage{amsmath}
\usepackage{geometry}
 \usepackage{caption}
\usepackage{bm}

\newdefinition{rmk}{Remark}
\newproof{pf}{Proof}
\newproof{pot}{Proof of Theorem \ref{thm2}}
\newcommand{\be}{\begin{equation}}
\newcommand{\ee}{\end{equation}}
\newcommand{\bea}{\begin{eqnarray}}
\newcommand{\eea}{\end{eqnarray}}

\usepackage{graphicx}% Include figure files
\begin{document}

\begin{frontmatter}
\title{  Chiral symmetry and Heun's equation}
\author{Yoon-Seok Choun}
\ead{ychoun@gmail.com}
\author{Sang-Jin Sin}
\ead{sjsin@hanyang.ac.kr}
\address{Department of Physics, Hanyang University, Seoul 04763, Korea }
\begin{abstract}
We show that the current quark mass  should vanish  to be consistent with the QCD color confinement: a bag model leads us to Heun's equation, which requests that not only the energy but also  the string tension should be quantized. This is due to the presence of higher order singularity which requests higher regularity condition demanding that parameters of the theory  should be related to one another.
As a result,  the Hadron spectrum is consistent with the Regge trajectory only when  quark mass vanishes.
Therefore, in this model,  the chiral symmetry is a consequence of the  confinement.  
\end{abstract}


\end{frontmatter}

\section{Introduction}

It has been understood that the QCD vacuum is working as a dual superconductor confining the color flux.
As a consequence the Hadron spectrum  is linear in quantum number $n$,
\be \alpha ' m^2=n+\beta,  \ee  which is the Regge trajectory  that led to the discovery of the string theory.
It is also known that chiral symmetry is one of the  leading   principle for the Hadron dynamics.  For the chiral symmetry, the mass of the quarks should vanish at least approximately.
Indeed, the current quark mass  contribute  less than 1\%  in counting the proton mass. 
However,   little is understood why this should be so.
In this paper, we  will relate the vanishingly small  quark mass to  the Regge trajectory itself,  which is a consequence of the confinement of the QCD color flux.

To show this,  we will use a bag model which will lead us
to the Heun's differential equation(DE), which can be characterized by a DE with more than three  singularities. The highest singularity at infinity and the one at 0, can be cancelled by   factoring out  two asymptotic   behaviors.
So if we have three singularities   the left over singularity leads us two term recurrence relation and we can make the wave function normalizable   by tuning  the energy parameter such that the remaining factor of the wave function is truncated to  a  polynomial, which is the well known energy quantization.

Now if we have four or  more singularities, then we need to tune two or more parameters of the differential equation to make the wave function normalizable.
In terms of the Schr\"odinger equation, the result is rather dramatic: Not only the energy but also  a parameter of the potential   must be quantized.
Sometimes, such extra quantization leads to obvious mismatch with the experimental data or well known principle unless  some parameter vanishes. 
In  our case,  the spectrum of the hadron will be
 consistent with Regge trajectory only when the current quark mass vanishes.
This  result can be used to relate the
origin of the chiral symmetry  to  QCD color confinement.

\section{Quark-antiquark  system with scalar interaction  }
To discuss the relation between the quark-mass and confinement, we use an old but simplest model where the confinement  dynamics is captured as a Regge trajectory. 
Lichtenberg and collaborators\cite{Lich1982} found a semi-relativistic Hamiltonian  which leads to a Krolikowski type second order differential equation \cite{Krol1980,Krol1981,Todo1971} in order to calculate meson and baryon masses in 1982.
In the center-of-mass system, the relativistic expression for the total energy $H$ of two free particles of masses $m_1$, and $m_2$, and three-momentum $\vec{\mathbf{p}}$ is
\begin{equation}
H = \sqrt{\vec{\mathbf{p}}^2 +m_1^2}+ \sqrt{\vec{\mathbf{p}}^2 +m_2^2}
\label{eq:ys}
\end{equation}
Let $S$ be an interaction which is a Lorentz scalar and $V$ be an interaction which  is a time component of a  Lorentz vector. Then it is natural to incorporate the $V$ and $S$ into (\ref{eq:ys}) by making the replacements
\begin{equation}
H\rightarrow H-V, \hspace{1cm} m_i \rightarrow m_i+\frac{1}{2}S, %\hspace{1cm} m_2 \rightarrow m_2+\frac{1}{2}S
\qquad i=1,2. \label{replace}
\end{equation}
Setting $m_{1}=m_{2}=m$, $V=0$    followed by (\ref{replace}),
and introducing the scalar potential $S=br$,
 G\"{u}rsey \textit{et al.}  got a spin-free Hamiltonian  for  the meson ($q\bar{q}$) system \cite{1985,1988,1991,2011}:
\begin{footnotesize}
\begin{equation}
H^2 = 4\left[(m+ \frac{1}{2}br)^2 + P_r^2 + \frac{L(L+1)}{r^2}\right]
\label{eq:67}
\end{equation}
\end{footnotesize}
where  we used  $\vec{\mathbf{p}}^2=  P_r^2 + \frac{L(L+1)}{r^2}$ with $P_r^2 = - \frac{\partial ^2}{\partial r^2} - \frac{2}{r} \frac{\partial}{\partial r} $, and $L$ is the angular momentum and $b$ is real positive constant.
Notice that the meaning of the linear scalar potential is to enforce the confinement of the quarks bound by a QCD flux string with constant string tension $b$. We will call the model  simply a bag model afterward.
 For  $m=0$, they could solve the eigenvalue problem  $H^{2}\Psi=E^{2}\Psi $  and obtained  energy eigenvalues
 \be
 E^2 = 4b\left(L+ N_r  +3/2\right), \label{energy0} \ee
 where $N_r=0,1,2,\cdots$ is the   quantum number counting  the  radial nodes.

 Notice that the energy is measured in the center of mass system therefore it is equal to the total mass of the system, namely the meson mass. Therefore above result is  consistent with  the Regge trajectories of slope $\frac{1}{4b}$. The purpose of this paper is to  understand  what happens in the case   $m\neq 0$.

\section{Heun's equation}
We start from the Schroedinger type equation
 $
H^{2}\Psi=E^{2}\Psi $ with $H^{2}$ given by the Eq.(\ref{eq:67}),
which can be considered as a non-relativistic Shcr\"odinger equation of the harmonic oscillator with  extra linear potential apart from the usual quadratic potential.

Factoring out the asymptotic behaviors of wave function $\Psi$ near $r=0$ and $r=\infty$ by
\be
{\footnotesize \Psi (r) = \exp\left(-\frac{b}{4}\left(r+\frac{2m}{b}\right)^2\right) r^L y(r)Y_L^{M}(\theta ,\phi)},\ee
  the differential equation for  (\ref{eq:67})  becomes
\begin{footnotesize}
\begin{equation}
r\frac{\partial^2{y}}{\partial{r}^2} + \left( - b r^2 -2m r +2(l+1)\right) \frac{\partial{y}}{\partial{r}} + \left( \left(\frac{E^2}{4}-
b\left(L+\frac{3}{2}\right)\right) r  -2m(L+1)\right) y = 0
\label{eq:BM}
\end{equation}
\end{footnotesize}
which is
%The radial wave functions satisfy the equation
%\begin{footnotesize}
%\begin{equation}
%\left[-\frac{\hbar ^2}{2m}  \left( \frac{d^2}{d{r}^2} +\frac{2}{r}\frac{d}{dr}\right) +\frac{\hbar ^2}{2m}\frac{L(L+1)}{r^2}   + V(r)\right] R(r) = E
%R(r)
%\label{qq:1}
%\end{equation}
%\end{footnotesize}
%with $V(r) = c r^2 + b r -a/r$ where $0\leq r<\infty $, $E$ is the eigenvalue and $L$ is the rotational quantum number: If $c\rightarrow b^2$, $b\rightarrow 4mb$, $a\rightarrow0$ and $E\rightarrow E^2 -4m^2$, it turns to be (\ref{eq:67}); (\ref{eq:67}) is the special case of (\ref{qq:1}).
%If one requires solutions in the form $ R(r) =r^L \exp\left( -\frac{\tilde{c}}{2}\tilde{r}^2-\frac{\tilde{b}}{2\tilde{c}}\tilde{r}\right)y(\rho)$
%where $r=\alpha \tilde{r} $ and $\rho = \sqrt{\tilde{c}}\tilde{r}$ with  $\tilde{E} = \alpha E $, $\tilde{c}^2= \alpha^3 c $, $\tilde{b}= \alpha^2 b $, $\tilde{a}= a $ and $\alpha = \frac{\hbar ^2}{2m}$, (\ref{qq:1}) becomes
%\begin{footnotesize}
%\begin{multline}
% \rho \frac{d^2 y(\rho)}{d\rho^2}  + \left(-2\rho^2 -\frac{\tilde{b}}{\tilde{c}^{3/2}}\rho + 2(L+1) \right) \frac{d y(\rho)}{d\rho} \\+ \left(
% \frac{1}{\tilde{c}}\left(\tilde{E} + \frac{\tilde{b}^2}{4\tilde{c}^2}-(2L+3)\tilde{c}\right) \rho +  \frac{\tilde{a}}{\sqrt{\tilde{c}}}
% -\frac{\tilde{b}(L+1)}{\tilde{c}^{3/2}} \right) y(\rho) = 0
% \label{qq:6}
%\end{multline}
%\end{footnotesize}
a   bi-confluent Heun (BCH) equation  whose canonical form is defined by
%\begin{footnotesize}
\begin{equation}
\rho\frac{d^2{y}}{d{\rho}^2} + \left( \mu \rho^2 + \varepsilon \rho + \nu  \right) \frac{d{y}}{d{\rho}} + \left( \Omega\rho + \varepsilon \omega \right) y = 0
\label{eq:BCH}
\end{equation}
%\end{footnotesize}
where $\mu$, $\varepsilon$, $\nu $, $\Omega$ and $\omega$ are real or imaginary
parameters. It has a regular singularity at the origin and an irregular singularity at the infinity of rank 2 \cite{NIST,Slavy2000}.

Substituting $y(\rho)= \sum_{n=0}^{\infty } d_n \rho^{n}$ into (\ref{eq:BCH}), we obtain the following recurrence relation:
\begin{equation}
d_{n+1}=A_n \;d_n +B_n \;d_{n-1} \hspace{1cm} \quad \hbox{for  } n \geq 1,
\label{eq:3}
\end{equation}
\be
\hbox{\!\!where   } A_n=-\frac{\varepsilon (n+\omega  )}{(n+1 )(n+\nu )},  \quad   B_n=-\frac{\Omega +\mu (n-1 )}{(n+1 )(n+\nu )},  \ee
and    $d_1= A_0 d_0$   for $ n=0$.
Comparing (\ref{eq:BM}) with (\ref{eq:BCH}),  the former is a special case of the latter with
 $\mu = -b$, $\varepsilon =  -2m $,
$\nu =  2(L+1)$, $\omega   = L+1 $ and
\be
\Omega  =E^{2}/4 -b(L+3/2). \label{Omega}
\ee
Unless  $y(\rho)$ is a polynomial, $\Psi$ is divergent as $ \rho\rightarrow \infty$.

\section{Normalizable solutions for the modified BCH equation}
It has been believed that we can make the   wave function normalizable whatever  form is the Schroeding equation by tuning the energy eigenvalue.
 However, what we shall meet is the fact that we need  to fine tune one more   parameter apart from the energy  in order to build normalizable (polynomial) solution for the Heun equations.
 This is because their series expansions consist of a three term recurrence relation given in Eq.(\ref{eq:3}) even after we factored out asymptotic behavior. Notice,
 on the other hand, hypergeometric-type functions gives only two term recursive relations, in which case we can  construct normalizable polynomial solution by tuning the single parameter, energy.
 Actually, the necessary and sufficient condition for constructing polynomials with a single parameter is that its power series should be reduced to the two term recurrence relation. For the Heun equation case, we cannot reduce its recursive relation to the two term case.
 We can build polynomials by fine tuning two parameters,   for example, $b$ and $E^{2}$.

For polynomials of (\ref{eq:BM}) around $r =0$, we treat $m$  as a free variable; consider $-\Omega /\mu= E^{2}/4b -(L+3/2) $ to be a positive integer; and treat $b$ as a fixed value.
Through (\ref{eq:3}), we are able to see  that a series expansion becomes a polynomial of degree $N$ if we  impose two conditions
\begin{equation}
B_{N+1}= d_{N+1}=0\hspace{1cm} \hbox{ for some }\;N\in \mathbb{N}_{0}
 \label{eq:21}
\end{equation}
Eq. (\ref{eq:21}) is sufficient to give  $d_{N+2}=d_{N+3}=d_{N+4}=\cdots=0$ successively and the solution to eq.(\ref{eq:BM}) becomes a polynomial of order $N$.

 To see what is going on we follow a few low order process. \\
For $N=0$,  Eq.(\ref{eq:21}) gives $B_1=\frac{-\Omega}{2(2L+3)}=0$ and $d_1=A_0 d_0= m d_0=0$.  If we choose $d_{0}=0$ the  whole series solution
vanishes.  Therefore there is no solution unless $m=0$, in which case  the solution is reduced to that of the Hypergeometric case  with $E^{2} = 4b(L+3/2) $.
%Here already, non-triviality of the
%solution request a special value of a  parameter of the   equation  other than the quantized energy. 
Since we are considering the case $m\neq 0$, we conclude  that there is no solution with radial nodal number $N=0$.

For $N=1$,
 $B_2=\frac{-\Omega +b}{3(2L+4)}$ and $d_2=A_1 d_1 + B_1 d_0 = (A_0 A_1 +B_1 ) d_0 = \left( \frac{4m^2 (L+1)(L+2)}{2(2L+2)(2L+3)} -\frac{b}{2(2L+3)}\right) d_0$. Requesting  both $B_{2}$ and $d_{2}$ to be zero, we get   $b=2m^2(L+2)$  and $E^{2} = 4b (L+1+3/2)=8m^{2}(L+2)(L+5/2)$ with $L=0,1,2,\cdots$.
In this case, $y(\rho)=\sum_{n=0}^{1}d_n \rho^n = 1+m \rho$  where  $d_0=1$ chosen for simplicity from now on.  Since N=0 is not allowed for $m\neq 0$, $N=1$ is the case containing the ground state.

For $N=2$, we have $B_3=\frac{-\Omega +2b}{4(2L+5)}$ and
 $d_3=A_2 d_2 + B_2 d_1 %= (A_0 A_1 A_2 + A_2 B_1 +A_0 B_2 ) d_0
 = \left( \frac{2(L+2)(L+3)m^3}{3(2L+3)(2L+4)}- \frac{2(L+3) m b}{3(2L+3)(2L+4)} -\frac{m b}{3(2L+4)}\right) d_0$. So, the  Eq.(\ref{eq:21})   gives
 $b=\frac{2m^2(L+2)(L+3)}{4L+9}$ and
  $E^{2} =
  \frac{8m^2(L+2)(L+3)(L+7/2)}{4L+9}  $  with $L=0,1,2,\cdots$.   Its eigenfunction is $y(\rho)=\sum_{n=0}^{2}d_n \rho^n = 1+m\rho +\frac{L+2}{4L+9}m^2 \rho^2$. %

For  larger $N$, the energy eigenvalue is determined from $B_{N+1}=0$, or equivalently $\Omega=-\mu N=bN$.  Eq.(\ref{Omega}) gives
\be
E^2 = 4 b \left( N +L+\frac{3}{2}\right), \label{energy}
\ee
 with  $L=0,1,2,\cdots, N$.   Allowed values of $b$'s are obtained from $d_{ N+1}=0$, which are quantized.  Its eigenfunction is  $N$-th order polynomial
\be y_{N}(\rho)=1+m\rho + \sum_{i=2}^{N}d_{i}\rho^{i} .\label{polynomial}
\ee

\section{Necessity of extra quantization}
We observed that both  $E$ and $b$ are quantized in order to have a polynomial  solution  (\ref{polynomial}) when we have three term recurrence relation.
However, for many people including the authors, it is not easy to accept the idea that  one more  parameter other than the energy should be quantized.
%None of the textbook states any restriction on potential  to have a well defined eigenvalue problem for the Hamiltonian.
The question of extra quantization is equivalent to asking whether   imposing both conditions in  Eq.(\ref{eq:21}) are the only way to get the normalizable solution, although it is clear that they are   sufficient.

 Here we demonstrate numerically that  we can not construct a normalizable solution  of the BCH equation  by tuning   only $E^2$ using the   shooting method.   %This will provide a complementary reasoning  to that based on the number of singularity of the differential equation provided in the precious section.
Let $m=1$ and $L=0$ in (\ref{eq:BM}) for simplicity.
According to previous section,  the ground state for $m\neq 1$  happened at $N=1$ with $E^2=E_0=40$,  but  $b=b_0=4$ was also required. In this case polynomial was given by $1+r$.
 What will happen if we do not request quantizing $b$?

Let  $b$ is  different  from the quantized value $b_{0}$ so that  $b = b_0+1.0$.  We look for a proper value of $E^2$ with  initial conditions  $y(0)=d_{0}=1, y'(0)=d_{1}=m=1$.
Then  we try to construct a normalizable solution by shooting method.

%numerical calculation shows that  its eigenfunction is not $1+r$ any more and we do not have a normalizable solution.

\begin{table}[!htb]
\tiny
    \begin{minipage}{.5\linewidth}
      \centering
        \begin{tabular}{|l|l|}
\hline
(1) & $E^2=E_0+7.496817$                         \\ \hline
(2) & $E^2=E_0+7.496818$                      \\ \hline
(3) & $E^2=E_0+7.49681789$           \\ \hline
(4) & $E^2=E_0+7.49681790$                  \\ \hline
(5) & $E^2=E_0+7.4968178907$                    \\ \hline
(6) & $E^2=E_0+7.4968178908$   \\ \hline
(7) & $E^2=E_0+7.496817890781$   \\ \hline
(8) & $E^2=E_0+7.496817890782$   \\ \hline
(9) & $E^2=E_0+7.4968178907817$   \\ \hline
(10) &$E^2=E_0+7.4968178907818$   \\ \hline
(11) & $E^2=E_0+7.49681789078176$   \\ \hline
(12) & $E^2=E_0+7.49681789078177$    \\ \hline
(13) & $E^2=E_0+7.496817890781766$    \\ \hline
(14) & $E^2=E_0+7.496817890781767$    \\ \hline
(15) & $E^2=E_0+7.4968178907817661$    \\ \hline
(16) & $E^2=E_0+7.4968178907817662$    \\ \hline
\end{tabular}
\caption{\scriptsize $E^2$ of $y(r)$ for $b = b_0+1.0$.\\ See  Fig.~\ref{Oneone}}.
    \end{minipage}%
    \begin{minipage}{.5\linewidth}
      \centering
        \begin{tabular}{|l|l|}
\hline
(1)  & $E^2=E_0-10^{-6}$  \\ \hline
(2)  & $E^2=E_0+10^{-6}$   \\ \hline
(3)  & $E^2=E_0-10^{-8}$ \\ \hline
(4)  & $E^2=E_0+10^{-8}$   \\ \hline
(5)  & $E^2=E_0-10^{-10}$   \\ \hline
(6)  & $E^2=E_0+10^{-10}$   \\ \hline
(7)  & $E^2=E_0-10^{-12}$  \\ \hline
(8)  & $E^2=E_0+10^{-12}$   \\ \hline
(9)  & $E^2=E_0-10^{-13}$  \\ \hline
(10) & $E^2=E_0+10^{-13}$   \\ \hline
(11) & $E^2=E_0-10^{-14}$  \\ \hline
(12) & $E^2=E_0+10^{-14}$   \\ \hline
(13) & $E^2=E_0-10^{-15}$  \\ \hline
(14) & $E^2=E_0+10^{-15}$  \\ \hline
(15) & $E^2=E_0-10^{-16}$  \\ \hline
(16) & $E^2=E_0+10^{-16}$  \\ \hline
\end{tabular}
 \caption{\scriptsize $E^2$ of $y(r)$ for $b = b_0$. \\See  Fig.~\ref{Twotwo}}.
    \end{minipage}
\end{table}

\begin{figure}[!htb]
\minipage{0.4\textwidth}
  \includegraphics[width=\linewidth]{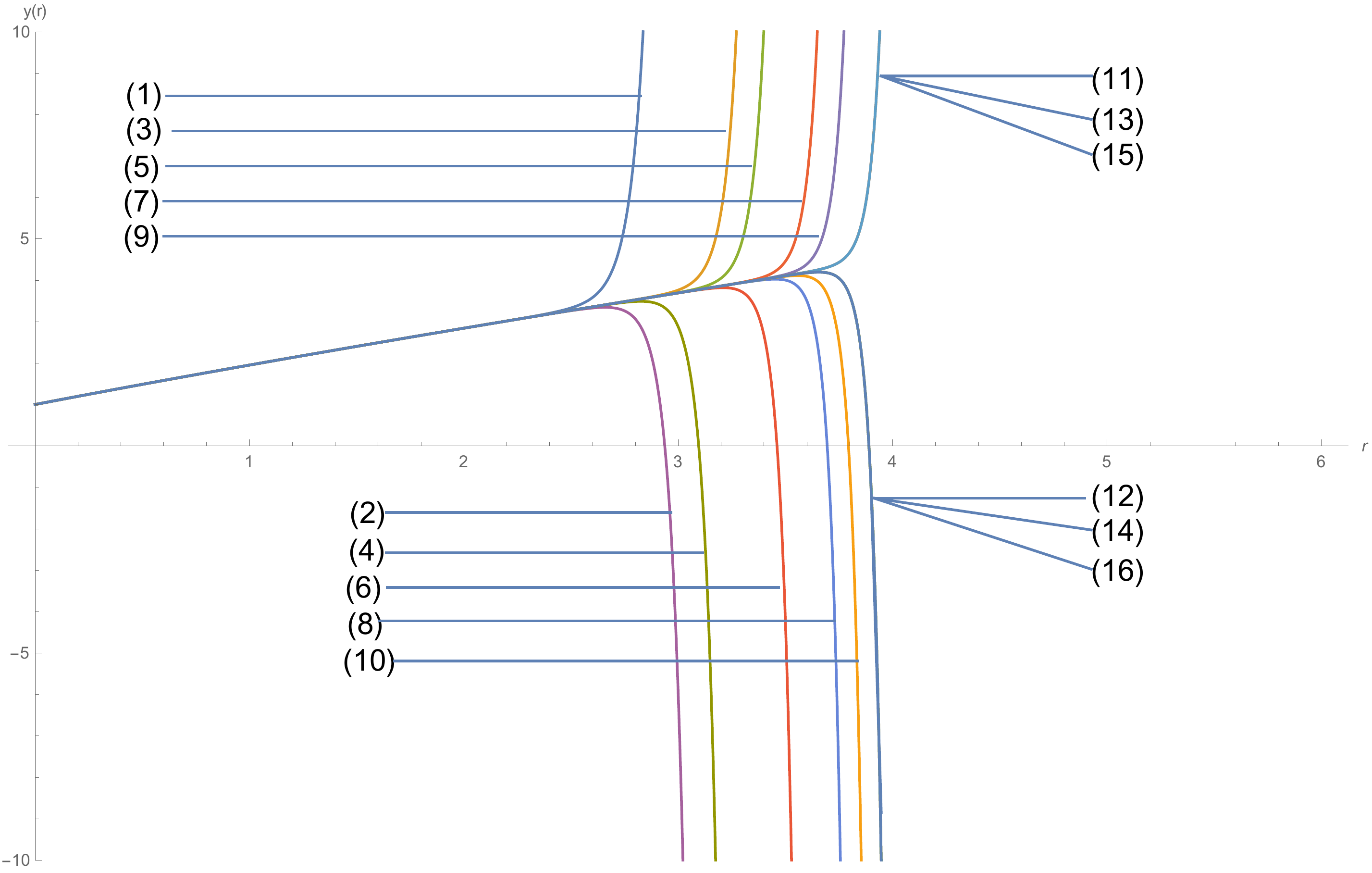}
  \caption{ $y(r)$ with a fixed $b=b_0+1.0$ and unfixed $E^2$'s  as  $m=1$}
\label{Oneone}
\endminipage\hfill
\minipage{0.4\textwidth}
 \includegraphics[width=\linewidth]{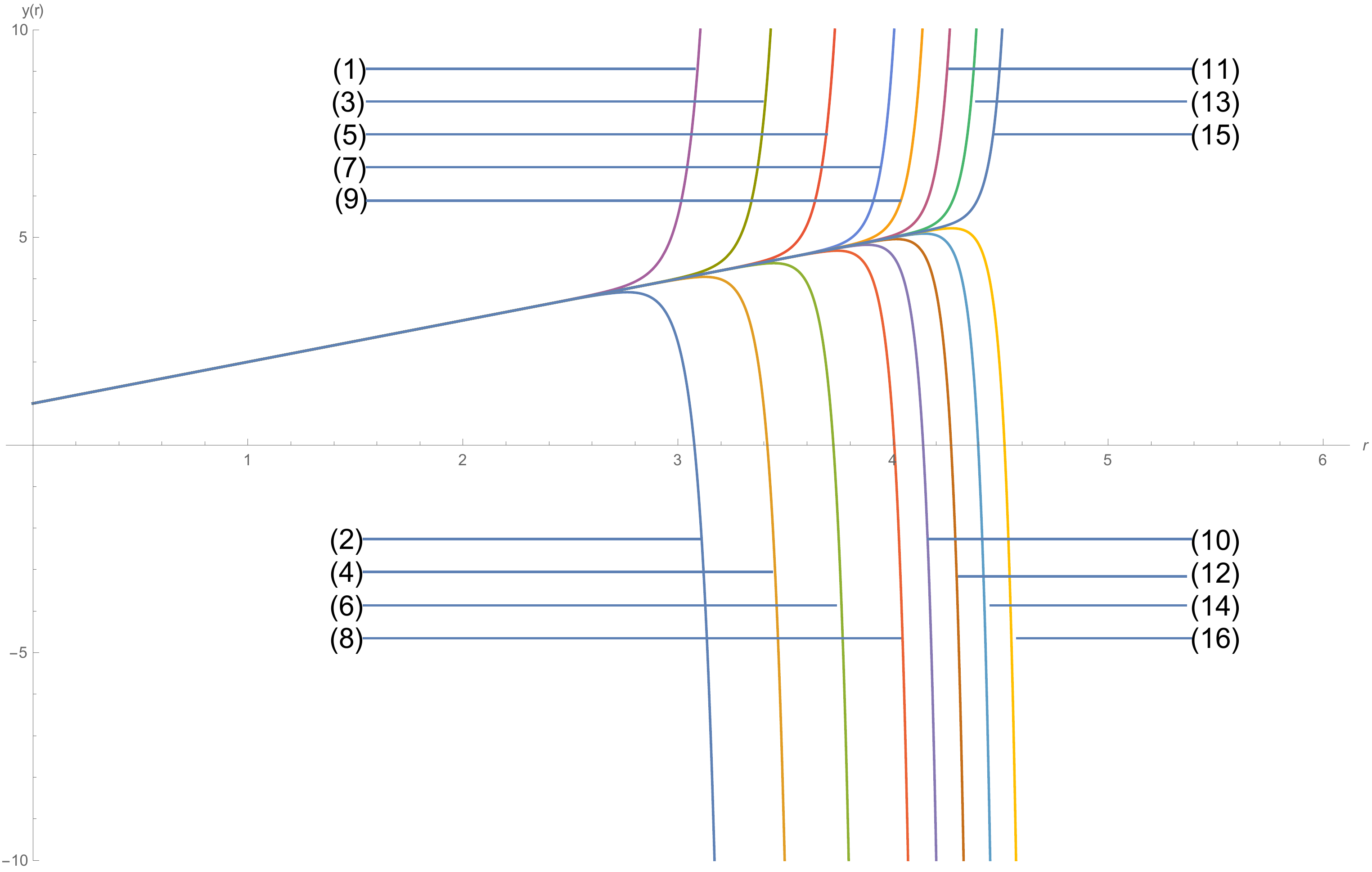}
  \caption{ $y(r)$ with a fixed $b=b_0$ and unfixed $E^2$'s as  $m=1$}
\label{Twotwo}
\endminipage
\end{figure}

In Fig.~\ref{Oneone} shows how the trial wave functions
approach to $1+r$ as we increase the precision of the
eigenvalue $E^{2}$. The odd numbered solutions (1),(2), ...
are undershooted ones and even numbered ones are overshooted ones. Starting from a under-shooted solution, one can increase the precision of the eigenvalue $E^{2}$ by increasing minimal amount in the next digit to get the over-shooted solution. Similarly, starting from a over-shooted solution one can increase the precision of the eigenvalue   by decreasing minimal amount in the next digit to get the under-shooted solution.
After a number of iterations,  the solutions stop to approach to $1+r$ although   we increase the precision by alternating the over- and under-shooting.  This can be seen from the Fig.~\ref{Oneone}:  there is a limit to pushing the solution  to the right as we see from overlapped solutions (11), (13), (15), and  (12), (14), (16).  When $E^2$ reaches around $E_0+7.49681789078176$, $y(r)$ starts to be flipped violently  without moving to the right any more.

This should be contrasted with $b = b_0$ case shown in
 Fig.~\ref{Twotwo} where the solution $y(r)$ is pushed to the right as $E^2$ approaches 40 with $b=b_0=4$ without problem.
 And we can  easily check that if $E^2 $ is exactly 40, $y(r)=1+r$ numerically also.

 Above demonstration  help us to accept necessity of  two quantized parameters ($E^2 $ and $b$) to create a polynomial,
when  a series solution of (\ref{eq:BM}) have of a three term recurrence relation.

%As $\tilde{a}=\tilde{b}=0$ in (\ref{qq:6}), its differential equation is a confluent hypergeometric equation which its series expansion is a two term recurrence relation. And $\tilde{c}$ is not a quantized value but a free parameter.
%Again, we are interested in a proper numerical value of $\widetilde{E}$ by applying the shooting method with Mathematica program. We choose $y'(10^{-50})=10^{-50}$ and $y(10^{-50})=1$ for initial conditions in (\ref{eq:41}).
%Fig.~\ref{Three} shows us that  $y(\xi)$  approaches the unity for the ground state with $\tilde{c}=c_0=2.5$ as $\widetilde{E}$ reaches to 7.5.
%In Fig.~\ref{Four}, we can observe that  $y(\xi)$  approaches the unity for the ground state for $\tilde{c}=c_0+1.0$ as $\widetilde{E}$ reaches to 10.5. These various examples tells us that we need two parameters ($\widetilde{E}$ and $\tilde{c}$) in order to make a polynomial of a three term recursive relation in a power series. On the other hand, we only need a single parameter ($\widetilde{E}$) to have a polynomial of a two term recursion relation in a series expansion.

\section{Quantization of $b$}  
Once we are convinced that both $E$ and $b$ are quantized,
we will find what are the available quantized values of $b$ for higher orders in the case of  a decrease  in energy level.  For  $N=L=10$, there are 5 possible real values of $b/m^2$: 0.366018, 0.579236, 1.03967, 2.35494 and 9.45702. We choose  the biggest real roots of $b/m^2$ in each case because it minimizes energy loss in a decrease  in energy level: Also, the biggest one makes $E_1<E_3<E_5<\cdots$ and $E_2<E_4<E_6<\cdots$, here, $E_i$ is the $i^{th}$ excited eigenvalue.

\begin{figure}[!htb]
\minipage{0.4\textwidth}
  \includegraphics[width=\linewidth]{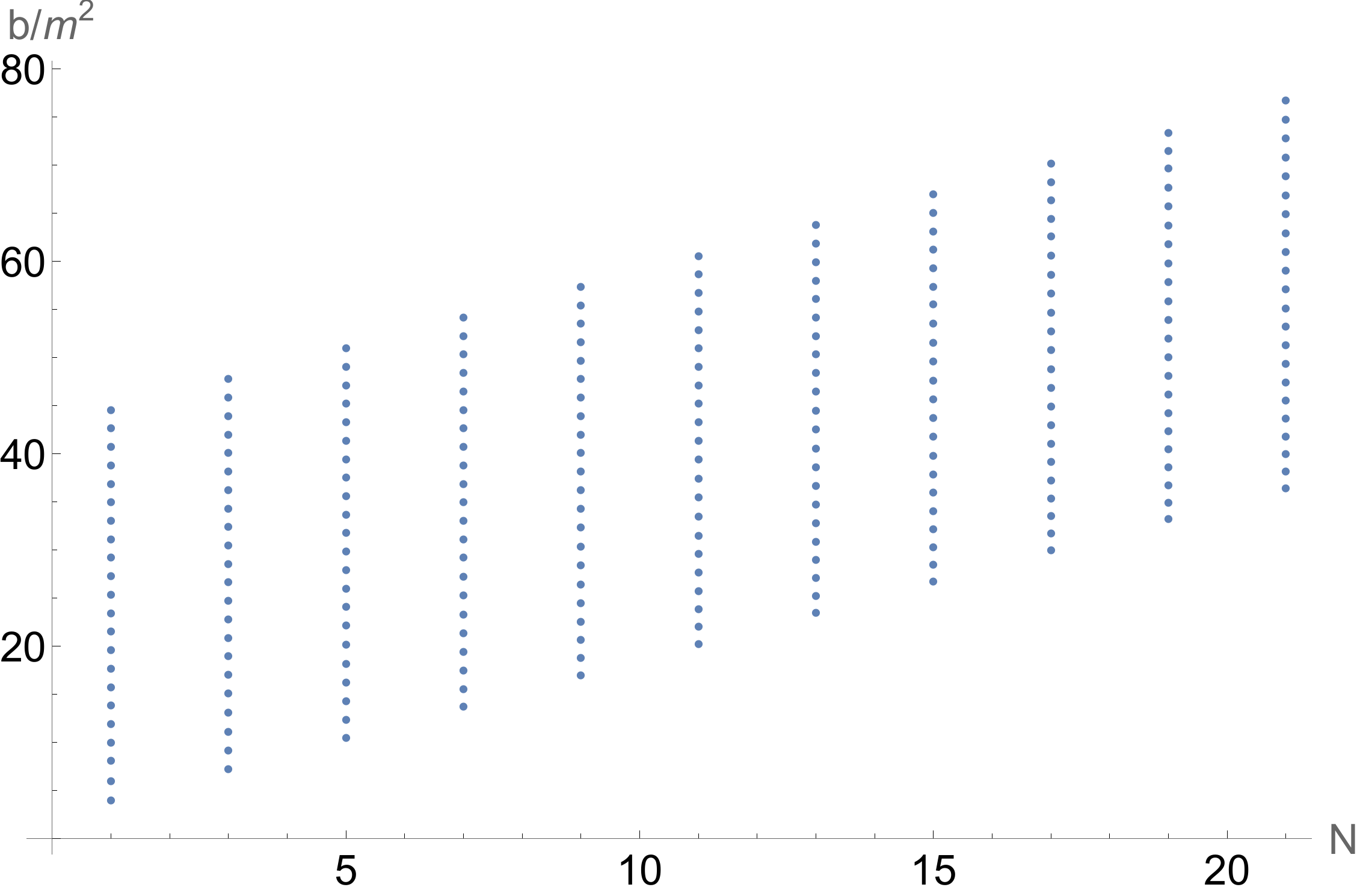}
  \caption{Real values of $b/m^2$'s for $N=1,3,5,\cdots$ \&  $L=0,1,2,\cdots, N$. }
\label{tension1}
\endminipage
\end{figure}

Fig.~\ref{tension1} shows us the biggest real values of $ b/m^2 $ with given odd $N$ and $L$. There are $(N+1)/2$ of $ b/m^2 $ corresponding to each $L=0, 1, 2, \cdots$.  
 In each  $ N $,  the lowest point is the numeric value of $ b/m^2 $ for $L=0$; the next point is  for  $L=1$;  the top point is for $L= 21 $.
We observe that  $b/m^2$ increases as $N$ increases with fixed $L$. As $N\rightarrow\infty$, 
 $b/m^2$ goes to infinity for any fixed $L$. Figs.~\ref{tension1} shows us that the gap between two successive points is constant with given $N$ as $L$ increases.
Similarly, we find that the biggest real values of $ b/m^2 $ with given even $N$ and $L$ is 1/4 less than the biggest real ones with odd $N$ and $L$.
%Similarly, Fig.~\ref{tension2} tells us the biggest real values of $ b/m^2 $ with given even $N$ and $L$.
%\begin{figure}[!htb]
%\minipage{0.4\textwidth}
% \includegraphics[width=\linewidth]{tension2.pdf}
%  \caption{Real values of $b/m^2$'s for $N=2,4,6,\cdots$ \&  $L=0,1,2,\cdots, N$.  }
%\label{tension2}
%\endminipage
%\end{figure}

\begin{figure}[!htb]
\minipage{0.4\textwidth}
  \includegraphics[width=\linewidth]{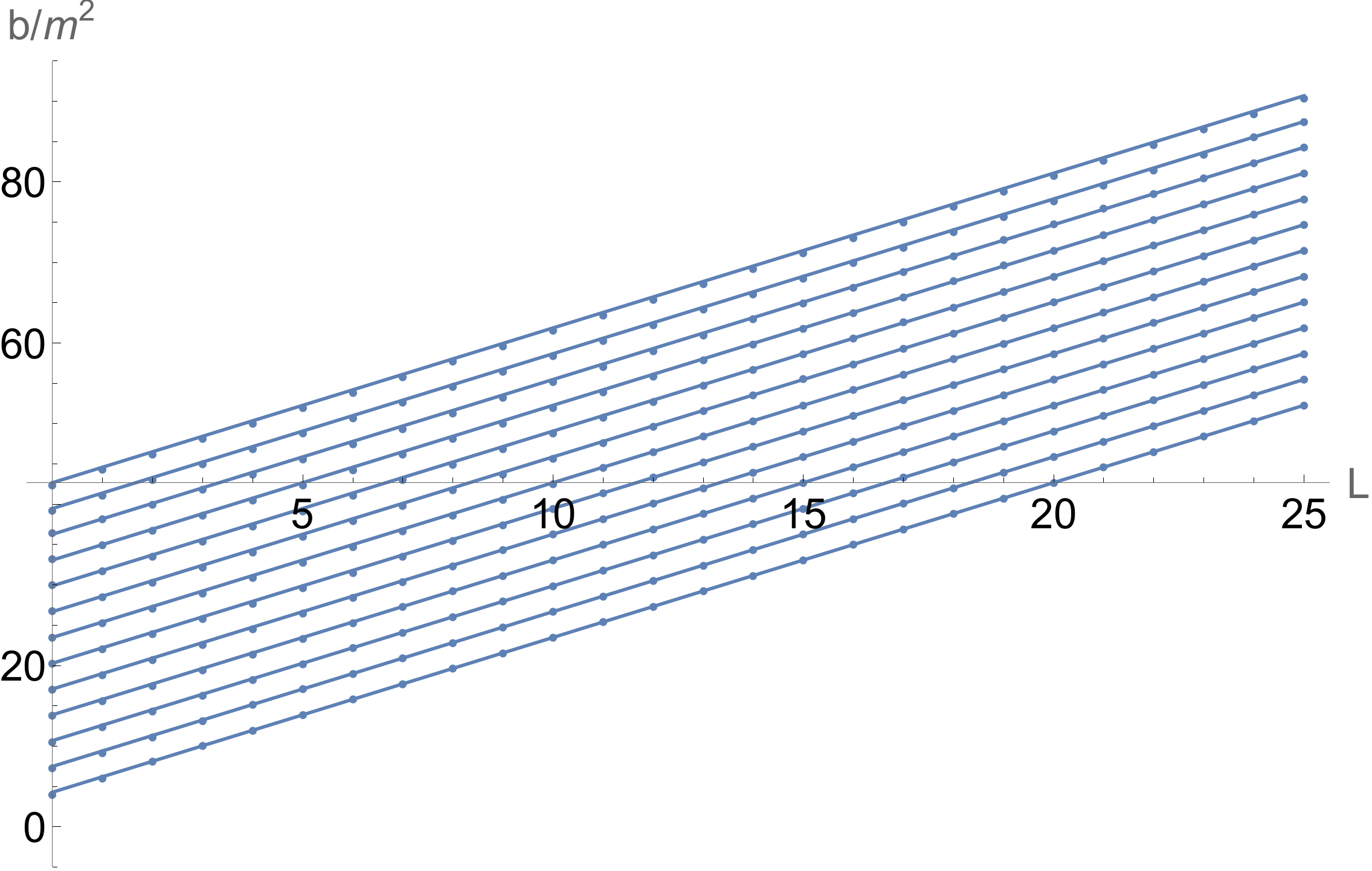}
  \caption{Fitting of $b/m^2$ by eq.(\ref{bm1}) as functions of $L$ with odd $N$.}
\label{odd1}
\endminipage
\end{figure}
\begin{figure}[!htb]
\minipage{0.4\textwidth}
  \includegraphics[width=\linewidth]{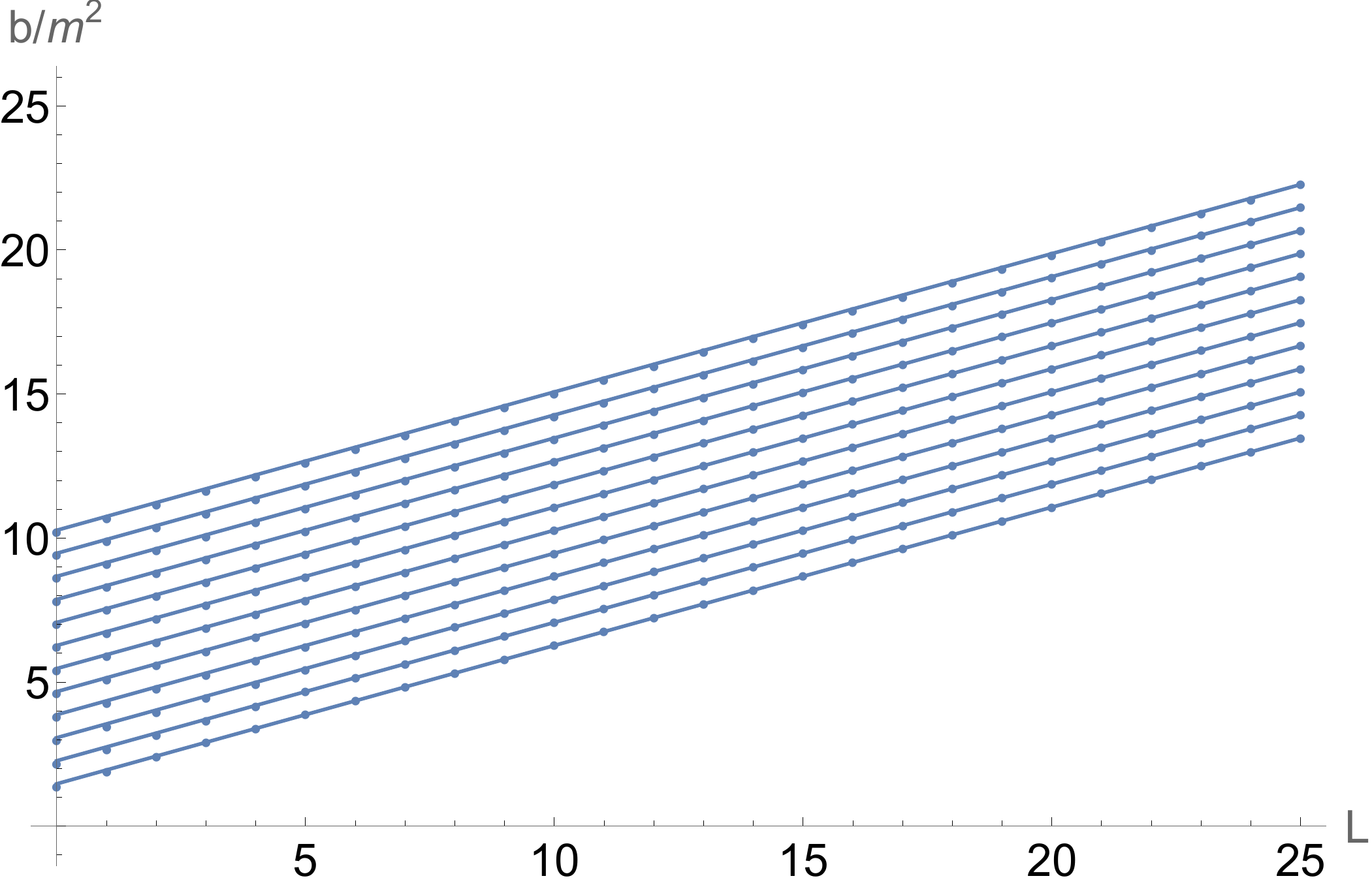}
  \caption{Fitting of $b/m^2$ by eq.(\ref{bm2}) as functions of $L$ with even $N$.}
\label{even1}
\endminipage
\end{figure}

% \begin{figure}[!htb]
% \minipage{0.4\textwidth}
%   \includegraphics[width=\linewidth]{odd1.pdf}
%   \caption{Fitting of $b/m^2$ by eq.(\ref{bm1}) as functions of $L$ with odd $N$.}
% \label{odd1}
% \endminipage
% \end{figure}
% \begin{figure}[!htb]
% \minipage{0.4\textwidth}
%  \includegraphics[width=\linewidth]{odd2.pdf}
%   \caption{Fitting of $b/m^2$ by eq.(\ref{bm1}) as functions of odd $N$ with $L$.}
% \label{odd2}
% \endminipage
% \end{figure}

%  
% \begin{figure}[!htb]
% \minipage{0.4\textwidth}
%   \includegraphics[width=\linewidth]{even1.pdf}
%   \caption{Fitting of $b/m^2$ by eq.(\ref{bm2}) as functions of $L$ with even $N$.}
% \label{even1}
% \endminipage
% \end{figure}
% \begin{figure}[!htb]
% \minipage{0.4\textwidth}
%  \includegraphics[width=\linewidth]{even2.pdf}
%   \caption{Fitting of $b/m^2$ by eq.(\ref{bm2}) as functions of even $N$ with $L$.}
% \label{even2}
% \endminipage
% \end{figure}

For odd $N$, Fig.~\ref{odd1} shows us that the allowed value of $b/m^{2}$ is linear  in $L$ and can be approximated by the rational function
 \be \frac{b}{m^2} = \frac{8}{5}\left( N+\frac{6}{5}L+\frac{5}{3}\right) \label{bm1}\ee . 
 For the figure, we calculated  625 different values of $ b/m^2 $'s  at various $(N,L)$. The lowest fit line is for $N=1$, the top one which has the most steep slop is for $N=25$. 
  \\
%Fig.~\ref{odd2} shows the result  as a function of  odd $N$   for  a few fixed $L$'s; the lowest fit line is for $L=0$ and the top one is for $L=23$.

For even $N$, Fig.~\ref{even1} shows us that the allowed value of $b/m^{2}$ is also linear  in $L$ and can be approximated 
 \be \frac{b}{m^2} = \frac{2}{5}\left( N+\frac{6}{5}L+\frac{5}{3}\right) \label{bm2}\ee . 
  \\
%Fig.~\ref{even2} shows the result  as a function of  even $N$   for  a few fixed $L$'s; the lowest fit line is for $L=0$ and the top one is for $L=22$.

By substituting  eq.(\ref{bm1}) and eq.(\ref{bm2}) into eq.(\ref{energy}),
we get the experimental fit to the eigenvalue $E^2$:

For odd $N$,
\be
  E^2 \approx  \frac{32}{5}\left( N+\frac{6}{5}L+\frac{5}{3}\right) \left( N +L+\frac{3}{2}\right)m^2 \label{energyF}
 \ee
 
For even $N$,
\be
   E^2 \approx  \frac{8}{5}\left( N+\frac{6}{5}L+\frac{5}{3}\right) \left( N +L+\frac{3}{2}\right)m^2 \label{energyG}
 \ee 

% Fig.~\ref{mesonN} shows  the fit lines of $E^2/m^2$ as function of   $N$  with a few given $L$'s.
% %It converges to 5 approximately as $N$ goes to infinity;
% The lowest fit line is for $L=0$ and  the top one is for $L=10$.
%Fig.~\ref{mesonL} shows  the fit lines of $E^2/m^2$ as functions of $L$ with a few given $N$'s; the lowest fit line is for $N=40$ and the top one is for $N=26$. It tells us that  $E^2/m^2 \propto L^2$ with given $N$.
%
%
%\begin{figure}[!htb]
%\minipage{0.4\textwidth}
%  \includegraphics[width=\linewidth]{mesonN.pdf}
%  \caption{Fit lines of $E^2/m^2$' as a function of $N$
%  %with given $L$'s as $K=0$
%\\   ($L=0,1,2,\cdots,10)$.
%   }
%\label{mesonN}
%\endminipage\hfill
%\minipage{0.4\textwidth}
% \includegraphics[width=\linewidth]{mesonL.pdf}
%  \caption{Fit lines of $E^2/m^2$as a function of  $L$ for    ($N=26,27,\cdots,40$)}
%\label{mesonL}
%\endminipage
%\end{figure}

One obvious consequence of our analysis is that the mass spectrum which is roughly given by Eq.(\ref{energyF}) and Eq.(\ref{energyG}) can not be linear in $N$  unlike  $m=0$ case given in Eq.(\ref{energy0}).
This is attributed to the fact that higher order singularity of the differential equation requests higher regularity condition so  that $b$ should be determined by other parameters,  which in turn introduces  extra dependence of  $E^{2}$   on $N$ and $L$  through that of $b$.

 \section{Conclusion}
 In this paper, we considered the spectrum of a bag model with
 non-zero quark mass, and found that
 the mass of the hadrons are non-linear, while it is linear if the quark mass is zero.
%  we have shown that there is a reason to attribute
% the vanishingly small mass of the light quarks to the confinement phenomena itself. Namely,
%  in the context of the model we considered,  if current quark mass  $m$ is non-zero,
%  the resulting spectrum becomes nonlinear, which is inconsistent with the well known hadron spectrum, the linear Regge trajectory.
 In the   model given by Eq.(\ref{eq:67}), the
 presence of current quark mass introduces a higher order singularity which requires extra regularity condition so that
  the string tension $b$ must  be
  related to the other parameter of the model and should be quantized.
   As a result,   $b$ gets extra $N$ dependence
and  the spectrum becomes non linear, which is inconsistent with the Regge trajectory that is tied with the color confinement.
In this sense and context, we can say that chiral symmetry is induced by  the color confinement. It would be interesting if similar argument can be done in other approach of hadrons.

 %\acknowledgments
\section*{Acknowledgements}
 We acknowledge the useful discussion with  Eunseok Oh.  This  work is supported by Mid-career Researcher Program through the National Research Foundation of Korea grant No. NRF-2016R1A2B3007687.
 %YS is  supported   by Basic Science Research Program through NRF grant No. NRF-2016R1D1A1B03931443.
%\end{acknowledgements}

%\bibliographystyle{}%{model1a-num-names}
%\bibliography{}%{<your-bib-database>}

\end{document}